\documentclass[prl,aps,twocolumn,amsfonts,showpacs,superscriptaddress,floatfix]{revtex4}

\usepackage{epsfig}
\usepackage{psfrag}
\usepackage{color}
\usepackage{graphicx}
\usepackage{subfigure}

\begin{document}


\title{A Renormalization Group For 
Treating 2D Coupled Arrays of Continuum 1D Systems}
\author{Robert M. Konik}
\affiliation{CMPMS Department, Brookhaven National Laboratory, Upton, NY 11973}
\author{Yury Adamov}
\affiliation{Department of Physics, Texas A\&M University, College Station, TX 77843}
\date{June 23, 2007}

\begin{abstract}
We study 
the spectrum of two dimensional coupled arrays
of continuum one-dimensional systems
by wedding a density matrix renormalization group procedure
to a renormalization group improved truncated spectrum approach.
To illustrate the approach we study the
spectrum of large arrays of coupled quantum
Ising chains.  We demonstrate explicitly
that the method can treat the
various regimes of chains, in particular
the three dimensional Ising ordering transition the chains undergo as a function
of interchain coupling.
\end{abstract}
\maketitle

The density matrix renormalization group (DMRG) \cite{white} is 
one of the primary theoretical tools for the quantitative description of 
low dimensional
lattice models.  For a wide range of one dimensional (1D) lattice models,
DMRG can characterize the model's spectrum and correlation functions \cite{kuhn}.
While there have been notable recent advancements \cite{moukouri,white2},
its use on 2D lattice models is more circumscribed \cite{hallberg}.

There exist several strategies to apply DMRG to 2D models.
In the first, the 2D lattice is reduced to a 1D
lattice with long range interactions \cite{noack}.  
A second approach sees short chains treated as individual lattice sites, allowing
the 1D DMRG algorithm to be applied to a model with short ranged interactions
directly \cite{jongh}.  In a 
more sophisticated variant of this methodology,
the DMRG is applied in a two stage process \cite{moukouri}.  The 2D
system is first divided into a set of coupled 1D chains and 
the DMRG is used to determine a low
energy reduction thereof.  For the
second stage, the reduced chains, coupled together and treated as individual
lattice sites in a 1D lattice, are analyzed again using DMRG.  In a final approach,
the 1D matrix product states underlying the DMRG algorithm \cite{ostlund} are replaced
by a higher dimensional generalization, projected entangled pair states \cite{ver}.

In this letter we present a distinct approach to applying DMRG to
2D models.  This approach trades upon a description of a
2D system as a mixture of continuum and discrete
degrees of freedom.  In particular, we approach 2D systems
as coupled arrays of continuum 1D chains with truncated Hilbert
spaces.  This methodology offers several
distinct advantages.  It allows us to treat
any 2D strongly correlated model provided it can be
conceived as composed of continuum 1D subunits.
Furthermore, the approach affords superior
finite size scaling.  As a function of the length, $R$, of the composite
1D systems, finite size corrections behave exponentially.  This implies that
we can access, at the very least, the infinite volume limit in the 
dimension parallel to the chains.
Finally, the truncation of the underlying 1D Hilbert
space dramatically lessens the numerical burden of the DMRG algorithm,
while providing a natural means to perform a Wilsonian renormalization group (RG)
improvement of any resulting answer \cite{first}.

The specific type of system that we propose to study takes the form,
\begin{equation}\label{ei}
H = \sum_i H^{1D}_i + J \sum_{<ij>}{\cal O}_i{\cal O}_j.
\end{equation}
The 1D continuum subunits of the array are governed by $H^{1D}_i$ which we insist 
must be either gapless (and so
governed by a conformal field theory) or gapped but integrable \cite{note}.  Thus this method
can study, for example, arrays of Luttinger liquids and a wide range of coupled 
1D Mott insulators.    The subunits are coupled together via
the nearest neighbour coupling $J{\cal O}_i{\cal O}_j$, where ${\cal O}_i$ is an
operator defined along the i-th continuum chain.  This coupling should
be relevant but can be of arbitrary strength. 

The analysis of the arrays proceeds in two conceptual steps.  In the first step we follow
Zamolodchikov's pioneering numerical analysis of perturbed gapless 1D continuum theories \cite{zamo},
an approach termed the truncated spectrum approach (TSA).
We thus place the 1D chains on a ring of circumference,
$R$.  Unlike DMRG applied to pure lattice models, periodic boundary conditions along the chains
can be employed without issue.  By working at finite $R$, we discretize the 1D spectrum.  
This permits the states in the chains' spectrum to 
be ordered in energy, i.e. $|1\rangle, |2\rangle,\ldots$, and then truncated at some finite
cutoff, $E_c$, leaving us with a finite number of states in the theory.  With these
alterations we nonetheless
remain in an excellent position to obtain information regarding the full theory in infinite volume.
In Ref. \cite{zamo}, a critical Ising chain in a magnetic field was studied.  Choosing $E_c$
so that a mere $39$ states were kept, infinite volume results were reproduced within an error
of $2\%$ (via diagonalization of a $39 \times 39$ matrix).   It was this finding of excellent results
at little numerical cost that motivated us to apply the TSA to more complicated situations \cite{first},
including, as here, arrays of 1D chains.

A part of the TSA's success is predicated on embedding non-perturbative information
into the initial computation.  In the context at hand this means that 
given any two states, $|k\rangle, |l\rangle$, we can {\it exactly} compute matrix elements of the
form $\langle k |{\cal O}_i | l\rangle$ by virtue
of $H^{1D}_i$ either being conformal or integrable.  We stress that computation of such matrix
elements is always a practical possibility either by exploiting the algebraic structures inherent
in a conformal field theory or the form factor bootstrap approach in the case of a integrable
field theory \cite{ffprog}.

Having prepared the chains by truncating their finite volume spectrum, we proceed to the
actual analysis of coupled arrays.  We perform the analysis using the finite volume
DMRG algorithm.  Here each chain with truncated spectrum is treated as
if a single site.   We however adapt the DMRG algorithm to take into account that we are working
with a truncated spectrum.  This algorithm is outlined in Table 1.

\begin{table}
\begin{center}
\caption{Finite system DMRG algorithm adapted to the presence of a truncated spectrum.}
\begin{tabular}{ll}
\hline\hline
1. & Form initial Hamiltonian, $H_{m-1}$, (and any other needed \\
& operators, ${\cal O}_{m-1}$) of system block, $B_{m-1}$, of m-1 chains. \\
2. & Form Hamiltonian of superblock, $B_{m-1}\bullet\bullet B_{m-1}$, of 2m \\
& chains, only keeping states whose energy, governed by \\
& $H_{m-1} \otimes H^{1D} \otimes H^{1D} \otimes H_{m-1}$, i.e. coupling between \\
& the blocks is absent, is less than $E_c$.\\
3. & Find low-energy target state(s) of superblock. Form \\
& reduced density matrix of m-chain block, $\rho_m$. \\
4. & Diagonalize $\rho_m$, keeping sufficient eigenstates to\\
& obtain a truncation error of less than $\epsilon_c$.\\
5. & Recast $H_m$ and ${\cal O}_m$ in basis of kept eigenstates
of $\rho_m$,\\
& obtaining $\bar{H}_m$ and $\bar{{\cal O}}_m$.\\
6. & Diagonalize $\bar{H}_m$, obtaining $\bar{H}^*_m$.
Rewrite $\bar{{\cal O}}_m$ in terms of\\
& eigenstates of $\bar{H}^*_m$.\\
7. & Form new superblock, $B_{m}\bullet\bullet B_{m}$, of length 2m+2. \\
& As determined by ${\bar H}^*_m \otimes H^{1D} \otimes H^{1D} \otimes {\bar H}^*_m$, only keep\\
& states whose energy is less than $E_c$.\\
8. & Repeat 3-7 with $m\rightarrow m+1$ until desired length of\\
&system is obtained.\\
9. & Perform finite volume sweeps until convergence.\\
\hline\hline
\end{tabular}
\end{center}
\vskip -.1in
\end{table}

Most steps of our finite system DMRG algorithm are unchanged from the standard algorithm \cite{white1}.
One primary difference lies in the formation of the superblock Hilbert space and Hamiltonian (steps 2 and 7
in Table 1).  
Instead of keeping all states in the Hilbert space, we insist that the state have 
an energy less than the truncation energy, $E_c$, as governed by the Hamiltonian
of the uncoupled blocks.  Concomitantly in order to meaningfully associate
an energy with an arbitrary state of a superblock, we must
be able to assign an energy to the states of each block.  

\begin{figure}[tbh]
\includegraphics[height=1.5in,width=2.in,angle=0]{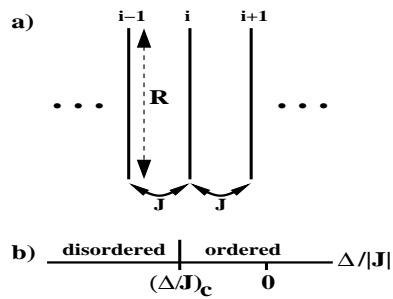}
\caption{a) A sketch of an array of Ising chains, each of length R; b)
a zero temperature phase diagram of the coupled chains showing both
the ordered and disordered regions.}
\end{figure}

This mandates that at each iteration of the DMRG we must
not only recast the block Hamiltonian, $H_m$, in terms of the kept eigenstates of the reduced
density matrix, $\rho_m$ (step 5), obtaining $\bar{H}_m$, but we must diagonalize $\bar{H}_m$ (step 6).
Working then with a basis of eigenstates of $\bar{H}_m$ allows us to associate a definite
energy to the superblock states formed in the next step (step 7).

A notable operational feature of this DMRG is the relatively small number of states we need
to keep from the diagonalization of the reduced density matrix in order to achieve a given truncation
error.  In the example of arrays of coupled quantum Ising chains that we consider below, the
number of kept states needed to achieve truncation errors on the order of $5\times 10^{-5}$
ranges from $10-40$ deep in the ordered phase of the chains to $50-90$ near the critical
value of the interchain coupling where the chains order.

The relatively small number of states needed to achieve a given accuracy mimics Zamolodchikov's
original finding that a relatively small number of states was needed to describe a critical
Ising chain in a magnetic field.  But it also reflects the presence of a truncated spectrum
in our approach.  As was shown in Ref. \cite{callan}, the entanglement entropy that arises
from dividing a 2D system into two behaves as the cutoff, $E_c$.  As the entanglement
entropy is one measure of whether a DMRG-like algorithm will be successful \cite{ver,hallberg}, 
we believe that our introduction
of a cutoff into the problem is a key feature of our approach.

\begin{figure}[tbh]
\includegraphics[height=1.8in,width=2.9in,angle=0]{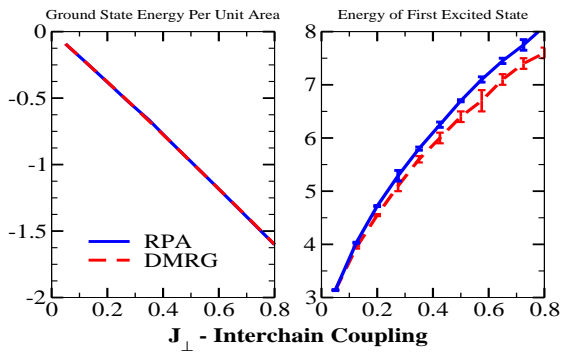}
\caption{The ground state and first excited state energy of
a 100 chain array as
a function of interchain coupling 
obtained with both a random phase approximation (blue solid) 
and our DMRG methodology (red dashed).  Here $\Delta = 1$.}
\end{figure}

A second key but related feature that marks our use of the DMRG algorithm is the use of RG 
improvement.  For sufficiently large truncation energies, $E_c$, the quantities of interest
(whether it be the spectrum or some observable) satisfy a one-loop RG flow \cite{first}:
\begin{equation}\label{eii}
\frac{d\Delta Q}{d\ln E_c} = -g\Delta Q.
\end{equation}
Here $\Delta Q = Q(E_c)-Q(E_c=\infty)$ is the deviation of some
quantity as a function of the truncation energy from its value at $E_c=+\infty$.
The function $g$ is a constant determinable from high energy perturbation theory
and is related to the anomalous dimension of $Q$.  When Q is an energy of some state,
$g=1$.  
Knowledge of this flow allows us to run the DMRG algorithm at several different truncation
energies and then extrapolate the resulting flow to $E_c = +\infty$, so removing
any residual effect of our use of a finite value for $E_c$.

\noindent {\bf Coupled Ising Chains}: To demonstrate this methodology we consider arrays of 
Ising chains (Fig. 1a) coupled via a nearest spin-spin perturbation:
\begin{equation}\label{eiii}
H = \sum_i H^{1D~Ising}_i - J_\perp \sum_{<ij>}\sigma^z_i\sigma^z_j,
\end{equation}
where the summations are over the chains in the array.  
The lattice form of the 1D Ising model is
\begin{equation}\label{eiv}
H^{1D~Ising} = -J\sum_i(\sigma^z_i\sigma^z_{i+1}+ (1+g)\sigma^x_i).
\end{equation}
Its continuum version is a Majorana fermion with gap, $\Delta = gJ$.
We place the chain on a ring of circumference, $R$.  The corresponding 
Hilbert space of the chain divides itself
into two sectors, termed Neveu-Schwarz (NS) and Ramond (R).  At T=0, the chain can be either ordered
or disordered.  In its disordered phase, $\Delta < 0$, the NS/R sectors of the chain permit only
even/odd numbers of free fermionic excitations.  In the ordered phase, $\Delta > 0$,  
the two sectors both permit only states with even numbers of fermions.  
The matrix elements of the spin operator, $\langle i|\sigma^z|j\rangle$, 
needed to carry out the DMRG procedure
can be found in Ref. \cite{fonseca}.  These matrix elements are only
non-vanishing if the states $|i\rangle$ and $|j\rangle$ lie in different sectors
of the theory.  Because we couple the chains together with nearest neighbour spin bilinears,
i.e. $\sigma^z_i\sigma^z_{i+1}$, the Hilbert space of the full theory possesses two sectors.
Any given state of the full theory has a tensor form $\otimes_i |k_i\rangle$ where
$|k_i\rangle$ is a state on the i-th chain.  These two sectors are distinguished by whether
an even or odd number of the $|k_i\rangle$ lie in the NS sector (or equivalently, the R sector)
of the individual chains.  That the Hilbert space possess different sectors is a generic feature
of continuum models when placed in finite volume and is not particular to the Ising chains at hand.

\begin{figure}[tbh]
\vskip .1in
\includegraphics[height=1.8in,width=2.6in,angle=0]{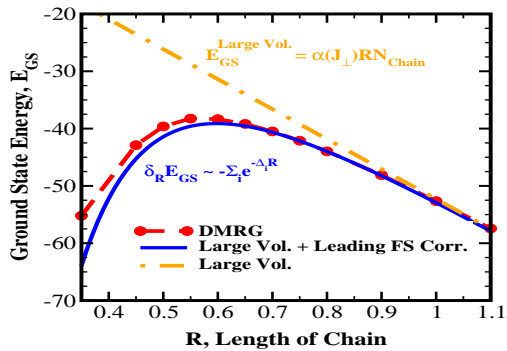}
\caption{The behavior of the ground state energy as
a function of chain length, $R$, for 100 chains.  $J_\perp = 0.275$.  In blue we see the analytic
prediction while in red we see the results of the DMRG.}
\end{figure}

We now submit the DMRG algorithm to a number of tests.
The first test that we apply relates to the behaviour of the spectrum of
the chains deep in their ordered phase ($\Delta > 0$ -- see the phase diagram in Figure 1b).  In this
region of the phase diagram, we expect a chain RPA analysis to be accurate \cite{carr}
and so provide
a baseline of comparison for our DMRG results.  The chain RPA analysis amounts to treating
the model
\begin{equation}\label{ev}
H^{I.C.Array}_{RPA} = -J\sum_i(\sigma^z_i\sigma^z_{i+1}+ (1+g)\sigma^x_i + h_{RPA}\sigma^z_i),
\end{equation}
where $h^{RPA}$ is chosen in the standard self-consistent fashion \cite{carr}.  We analyze
this model numerically using the truncated spectrum approach along the same lines
as Ref. \cite{fonseca}.  However akin to the discussion surrounding Eqn. (2) 
and in Ref. \cite{first}, we perform an RG
improvement of our numerical results.

In Fig. 2 we plot the ground 
and first excited state energy of a 100-chain array as a function of the interchain coupling
as computed by both the DMRG and the chain RPA analysis.  In order to optimize numerical performance,
we employ different chains lengths for different values of $J_\perp$.  To obtain
$R=\infty$ values for the gaps, $\Delta_{exc}$, of single excitations,
we can use any finite value of $R$
provided that we satisfy $R\Delta_{exc} \gg 1$.
Because finite $R$ corrections behave exponentially, $\delta_R \Delta_{exc} \sim e^{-R\Delta_{exc}}$, this
constraint need only be satisfied loosely.  In order 
to obtain a truncation error,$\epsilon_c$ 
of $1\times 10^{-4}$ we needed to keep at most 18 states.  Decreasing
the truncation error to $2.5\times 10^{-6}$ we needed to keep at most 34 states.
Moreover by comparing our results with an analysis of
50 coupled chains, we know that any finite size error related to studying only a finite number of chains
is extremely small.  The only significant uncertainty comes from the 
RG improvement -- applying Eqn. (2) --  and is the source of the error bars in Fig. 2.

\begin{figure}[tbh]
\vskip .1in
\includegraphics[height=1.8in,width=2.7in,angle=0]{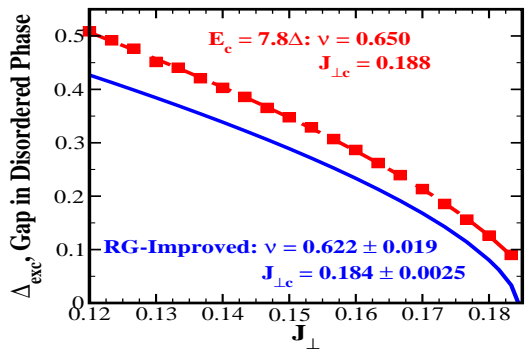}
\caption{The behavior of the gap in the disordered phase as a function of $J_\perp$
for 60 coupled chains at $R = 10\Delta$.
In red we plot the gap as determined for the truncation energy $E_c = 7.8\Delta$, while in blue we
plot the RG-improved curve.  For both, the values of $J_{\perp c}$ and $\nu$ are indicated.}
\end{figure}

We see from Fig. 2 that the ground state energy as determined by the DMRG agrees exceptionally well
with the chain RPA analysis even for values of the interchain coupling on the same
order as the gap.  The DMRG values of the gap for the first excited state however see increasing
deviations from the RPA values as the interchain coupling is increased.  These results
indicate that the truncated DMRG algorithm is operating as expected.

A more significant test of the algorithm is whether it predicts correctly the finite $R$ corrections.
These corrections can be computed analytically, at least at leading order.  According to Ref.
\cite {zamo}, they take the form
\begin{equation}\label{evi}
\delta_R E_{gs} = -\frac{1}{2\pi}\sum_b \sum_{k_y} \int dk_x e^{-R\epsilon_b(k_y,k_x)},
\end{equation}
and can be thought of as arising from spontaneous emission of excitations from the vacuum.
Here $\epsilon_b(k_y,k_x)$ is the dispersion of some band $b$
of excitations in the coupled chains.  $\delta_R E_{gs}$ sees contributions
from excitations with different (discrete) values of momenta, $k_y$, transverse to the chains as well as
excitations with different (continuum) values of the momenta, $k_x$, parallel to the chains.
While we cannot compute $\epsilon_b$ exactly, we can estimate it from the values of the
excited states as measured by the DMRG together with an RPA analysis \cite{carr}:
\begin{equation}\label{evii}
\epsilon_b(k_y,k_x) = (v_{Fx}^2k_x^2+\Delta_{b~exc}^2+2J_\perp Z_o(1-\cos(k_y)))^{1/2}.
\end{equation}
Here $\Delta_{b~exc}$ marks the lowest lying excitation in the band, while
$Z_o$ is the square of a matrix element of the spin operator
on a single chain, $Z_o = 2\langle 0|\sigma_i^z(0)|\epsilon_b(k_x=0,k_y=0)\rangle^2$.
The latter quantity 
is estimated from the same analysis used in computing the RPA energies of Fig. 1.

In Fig. 3 we plot $E_{gs}(R)$ for a particular value of interchain 
coupling ($J_\perp = 0.275$) using the correction in Eqn. 6 by taking into account
the two lowest energy bands in the theory (blue dashed line)
vs. the direct DMRG computation (red line).  We see that for all but the
smallest values of $R$ -- where our leading order analytic approximation breaks down -- the
DMRG and expected analytic values agree well.  For a point of comparison, we also plot the 
extrapolation of the energy from large values of $R$, where $E_{gs}$ scales linearly
with system volume, i.e. $E_{gs}(R) = \alpha(J_\perp) R N_{chain}$
(straight dashed orange curve).  For values of $R \sim \Delta_{b~exc}^{-1}$ the 
finite size corrections are exponentially suppressed, consistent with 
$\Delta_{b~exc} = 5.1\Delta$ for $J_\perp = 0.275$ (from Fig. 1).

The final test we put to the truncated DMRG algorithm is the chains' ordering
transition.  Beginning with chains in their disordered state ($\Delta < 0$) and
coupling them together with increasing $J_\perp$, they eventually order.  This transition
is in the same universality class as the 3D Ising model.  This implies that
the gap, $\Delta_{exc}$, in the disordered phase should vanish as 
$\Delta_{exc} \sim (J_{\perp c}-J_\perp)^\nu$
with $\nu = .630$ \cite{camp}.  From our DMRG analysis (see Fig. 4), we find after 
RG improvement,
$J_{\perp c} = .184\pm .0025$, together with good agreement for the critical exponent, $\nu = .622\pm .019$.
We see that RG improvement notably improves the results.  Computing instead $J_{\perp c}$ 
and $\nu$ from
the results obtained at the largest of the truncation energies employed 
($E_c = 7.8\Delta$), we obtain
$J_c = .1880$ and $\nu = 0.650$.

This DMRG approach to 2D arrays of continuum systems has a number of potentially valuable variations.
We first note that we have managed to treat 2D arrays using essentially a 1D DMRG algorithm.  
Ref. \cite{moukouri} has
demonstrated that the DMRG algorithm, if used in a two stage process, can treat systems of one
higher dimension.  This implies that it may well be possible to study {\it 3D arrays} using
our approach.  We also note that the algorithm we have outlined in this letter may see substantial
improvement if wedded to a numerical renormalization group (NRG) {\`a la Wilson} \cite{wilson}.  At
each DMRG iteration, one could perform a NRG akin to that described in Ref. \cite{first} to dramatically
increase the truncation energy being employed and so (hopefully) dramatically improve the results
of the procedure.

To summarize, we have demonstrated that the 1D DMRG algorithm can be directly applied to 2D arrays
of 1D continuum systems.   In particular we have shown this algorithm can describe the behavior of
large arrays of quantum Ising chains both in their ordered phase and in the vicinity of their 
order-disorder 
transition.  We expect that this procedure will produce quantitatively accurate results on a wide
variety of 2D systems in their infinite volume limit.

RMK and YA acknowledge support from the US DOE
(DE-AC02-98 CH 10886) together with useful discussions with A. Tsvelik.


\begin{references}
\bibitem{note} The requirement of integrability for gapped systems
can be relaxed but at additional numerical cost.
\end{references}
\end{document}